\documentclass {usjetpl}

\lat

\title{Magnetization loop and critical current of porous Bi-based HTSC}

\rtitle{Magnetization loop and critical current\ldots}

\author{D.M. Gokhfeld$^*$, D.A. Balaev$^*$, S.I Popkov$^{*+}$, K.A.
Shaykhutdinov$^*$, M.I. Petrov$^*$}

\rauthor{D.M. Gokhfeld, D.A. Balaev, S.I Popkov, K.A.
Shaykhutdinov, M.I. Petrov}

\address{$^*$L.V. Kirensky Institute of Physics SD
RAS, 660036, Krasnoyarsk, Russia\\
$^+$M.F. Reshetnev Siberian State Aerospace University, 660014,
Krasnoyarsk, Russia \\ E-mail: smp@iph.krasn.ru}

\abstract { The magnetization of porous
Bi$_{1.8}$Pb$_{0.3}$Sr$_{2}$Ca$_{2}$Cu$_{3}$O$_{x}$ have been
investigated. The experimental magnetization hysteretic loops of
$M(H)$ were described in the frames of Val'kov -- Khrustalev model
developed for type II granular superconductors.}

\PACS{ 74.25.Sv, 74.25.Ha, 74.72.Hs, 74.81.-g}

\begin{document}

\maketitle

\section{INTRODUCTION}

High-temperature superconductors (HTSs) having a foam structure
\cite{reddy,petr} is object of fundamental interest connecting
with study of transport and magnetic properties in disorder media
with fractal dimensions \cite{ftt05}. The high specific surface of
foam makes porous HTSs attractive for a practical application.
Influence of porosity of HTS on critical current is unclear. Only
article \cite{bart} concerning this matter can be referred where
critical state in superconducting single-crystalline
YBa$_{2}$Cu$_{3}$O$_{7}$ foam was considered.

\section{EXPERIMENTAL}

The standard ceramics method is employed to prepare the porous
Bi$_{1.8}$Pb$_{0.3}$Sr$_{2}$Ca$_{2}$Cu$_{3}$O$_{x}$ (BPSCCO),
except the final annealing. The time of synthesis was $\sim $400
h. The decomposition of calcium carbonate realizes during the
final annealing stage. The excess pressure of carbon dioxide
results in an increase of the material volume.

The density $\rho$ of the porous material equals 2.26 g/cm$^{3}$
(38 {\%} from theoretical one for BPSCCO). The SEM images show
that the porous BPSCCO has flakes-like microstructure formed by
the chaotic oriented crystallites \cite{petr,ftt05}. The
crystallites have thickness $\sim $2 $\mu$m and wide $\sim $10-20
$\mu$m.

To derive the effect of porosity on magnetic properties, bulk
BPSCCO was prepared from the initial porous BPSCCO by the standard
technique. For this dense HTS $\rho$ equals 95 {\%} from
theoretical one of BPSCCO.

The temperature of zero resistivity ($\le$ 1
$\mu\Omega$m$\cdot$cm) of porous and dense BPSCCO equals 107 K.

Measurements of magnetic field dependences of magnetization $M(H)$
in fields up to 60 kOe have been performed using the vibrating
sample magnetometer with the superconducting solenoid. The
specimens have the cylindrical form with height $\sim $4 mm and
diameter $\sim $0.7 mm.  The mass of sample is 0.0179 g for porous
BPSCCO and 0.0264 g for dense BPSCCO. The applied magnetic field
was parallel to axis of symmetry of cylinder sample.

\section{RESULTS AND DISCUSSION}

Figure 1 presents $M(H)$ dependences of porous BPSCCO with low
density and dense BPSCCO measured at liquid helium temperature. It
is clearly seen that the shape of this dependences is similar, but
the absolute value of diamagnetic response is higher in case of
porous superconducting material (in 2.4 times in units emu/g and
in 1.63 times in units emu).

In the Bean approximation to the critical state \cite{bean}, the
critical current densities can be estimated from the magnetization
data as \cite{frietz} $J_c = 30 \Delta M / 2R$, where $J_c$ is
measured in A/cm$^2$, $\Delta M = M^{+}$ - $M^{-}$ (in emu/cm$^3$)
is the width of the magnetization hysteresis loop at a certain
magnetic field, $R$ is the radius of cylinder sample (in cm). The
obtained zero field values of $J_c$ are 260 kA/cm$^2$ for porous
BPSCCO and in 1.63 times smaller, 160 kA/cm$^2$ for dense BPSCCO.
However given approach does not take into account granular
structure of HTSs. Application of a more appropriate model appears
to give some different results. The approach \cite{vkh} is
extension of the critical state theory for the case of granular
bulk HTS. The model \cite{vkh} allows to draw the hysteretic loops
in a broad magnetic field range. This model considers the
penetration of magnetic field in the crystallites having form of
cylinder. The general expression for performing $M(H)$ calculation
in \cite{vkh} is
\begin{equation}
4{\pi}M(H) = -H+(1-P){\mu}_nH+\frac{2}{R^2_{0}} \\ \int_0^{\infty}
{\varphi}(R) \: dR \int_0^R rB(r) \: dr ,
\end{equation}
where $P$ is the fraction of the superconductor concentrated in
the superconducting grains, ${\mu}_n$ is the magnetic permeability
of the intergranular material, ${\varphi}(R)$ is the distribution
density of the superconducting granules, $B(r)$ is the dependence
of the magnetic induction on the radius. The main fitting
parameter in the model \cite{vkh} is the average crystallite
radius $R$ and critical current density $J_{c}$, which determines
$B(r)$. A new form of dependence of $J_c$ on the magnetic
induction $B$ is employed in the model \cite{vkh} also. This
dependence is characterized by the presence of two magnetic
induction scales determined by the parameters $B_1$ and $B_0$.
These scales demarcate regions with different rates of decrease of
$J_c$. The dependence $J_c(B)$ has the following form \cite{vkh}:
\begin{equation}
J_c(B) = J_c(0)\left[\frac{1+Q
(B/B_1)}{1+B/B_1}+\left(\frac{B}{B_0}\right)^{\gamma}\right] ^{-1}
,
\end{equation}
where parameters $Q$ and $\gamma$ determine the rates of
variations of $J_c$ for the scales $B_1$ and $B_0$
correspondingly.

\begin{figure}[htbp]
\centerline{\includegraphics[width=6.18in,height=4.82in]{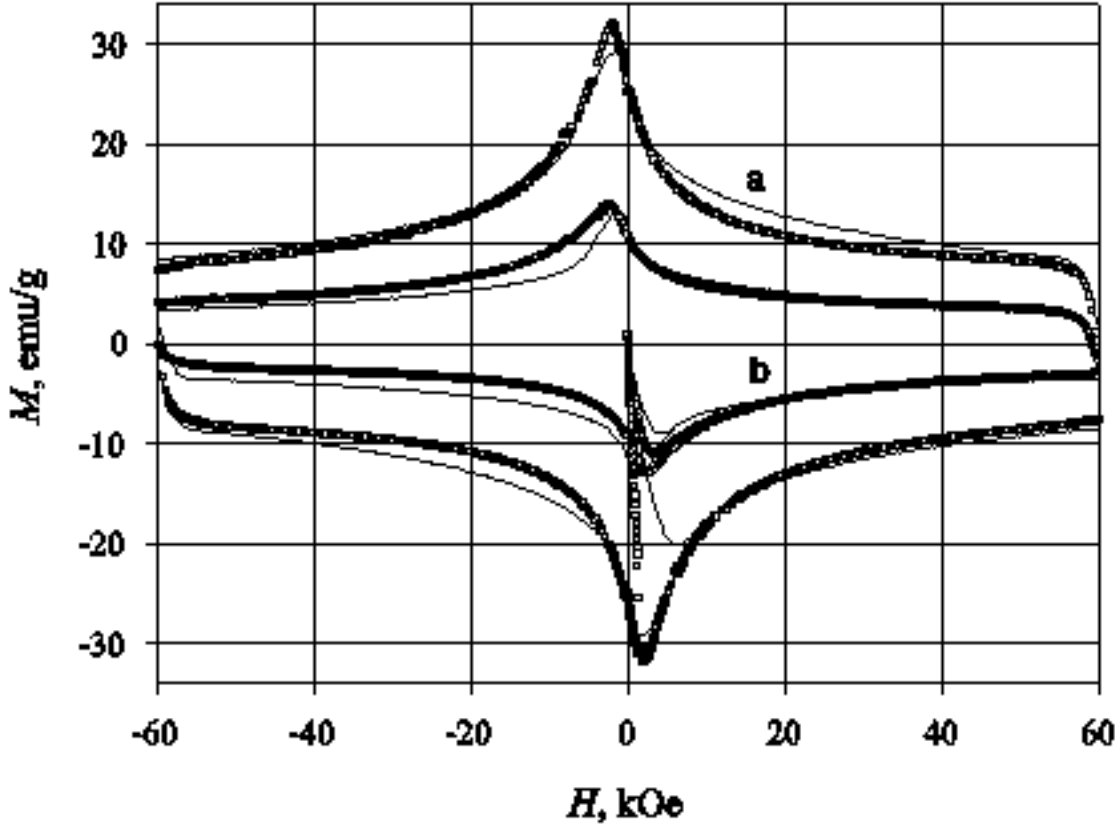}}
   \caption{Experimental $M(H)$ dependences of a) porous BPSCCO and
b) dense BPSCCO - points. Solid lines are results of computer
simulation of $M(H)$ dependence using theory \cite{vkh}.}
\label{fig1}
\end{figure}

The real crystallites of BPSCCO have a form of plates. Therefore
one should to modify the model \cite{vkh} to consider the
penetration of field in the randomly oriented flat crystallites.
It will be the task of future investigation. However we do not
expect a remarkable difference of magnetization picture.

The pores do not contribute to the magnetization of sample. We
accept $P$ equals to the density of material normalized on the
theoretical density of BPSCCO. Consequently $P$ = 0.38 for porous
BPSCCO and $P$ = 0.95 for dense BPSCCO. Two parameters are left in
the model: $J_c$ and $R$. The experimental magnetization curves
are described using the same value of $J_c$ = 2150 kA/cm$^2$ and
$R$ = 21 $\mu$m for porous BPSCCO and $R$ = 13 $\mu$m for dense
BPSCCO. The solid lines in fig. 1 are results of computer
simulation of $M(H)$ dependences in frames of model \cite{vkh}. A
satisfactory agreement between the experimental and theoretical
$M(H)$ curves is achieved. If we calculate $M(H)$ using lager
$J_c$ for case of porous BPSCCO than dense BPSCCO, the agreement
with the experiment becomes worse.

The grain boundaries limit critical current density of
polycrystalline HTSs \cite{hilg,phC,arxiv}. Especially remarkable
influence of grain boundaries are on the critical current density
obtained from the transport measurements $j_c$. But roles of grain
boundaries in porous HTSs are negligible \cite{petr,bart}. However
$j_c$ of porous BPSCCO is relatively small. Determination of $j_c$
by using criterion 1 $\mu$V cm gives 2-100 A/cm$^2$ at 4.2 K
\cite{ftt05}. These values are smaller than ones for bulk
polycrystalline BPSCCO ($\sim$ 100 A/cm$^2$). There are supported
to be two reasons of decreasing of $j_c$: the reduced effective
area of current flowing in porous HTS and the much smaller number
of percolation paths in the foam in comparison with the dense HTS.
We suppose that the high porosity of HTS does not lead to
enhancement of magnetic critical current density $J_c$ instead the
broader magnetization curve of porous BPSCCO than one of dense
BPSCCO. This suggestion is in contradiction with the prediction of
the Bean model. Thus the observed enhancement of the diamagnetic
response should be explained by other reasons except increasing of
$J_c$, e.g. larger size of the crystallites in porous BPSCCO.

\section{CONCLUSION}

We investigated and compared the magnetization curves of porous
BPSCCO and dense one. The experimental magnetization hysteretic
loops of $M(H)$ were described in the frames of Val'kov --
Khrustalev model \cite{vkh} developed for type II granular
superconductors. The increasing of magnetic critical current
density $J_c$ is not found in the porous BPSCCO. Contrary, growth
of the porosity is accompanied by decreasing of number of
percolation paths that reduces the transport critical current
density $j_c$. Discovered enhancement of the diamagnetic response
in porous HTSs is very attractive for practical applications, e.
g. superconductor bearings and levitators. \\

We are thankful to V.V. Val'kov and S.A. Satzuk for fruitful
discussions.

This work is supported by program of President of Russian
Federation for support of young scientists, grant MK 1682.2004.2.,
and by Krasnoyarsk Regional  Scientific Foundation (KRSF), grant
12F0033C


\begin{thebibliography}{10}

\bibitem{reddy} E.S. Reddy, G.J. Schmitz, Supercond. Sci. Technol. 15, 21 (2002).

\bibitem{petr} M.I. Petrov, T.N. Tetuyeva, L.I. Kveglis, A.A. Efremov, G.M. Zeer,
K.A. Shaihutdinov, D.A. Balaev, S.I. Popkov, S.G. Ovchinnikov,
Techn. Phys. Lett. 29, 986 (2003) [Pis'ma v Zh. Tekhn. Fiz., 29,
40 (2003)].

\bibitem{ftt05} D.A. Balaev, I.L. Belozerova, D.M. Gokhfeld, L.V. Kashkina, Yu.I. Kuzmin, C.R. Michel,
 M.I. Petrov, S.I. Popkov, K.A. Shaykhutdinov, Phys. Solid State (will be published).

\bibitem{bart} E. Bartolom\'{e}, X. Granados, T. Puig, X. Obradors, E.S. Reddy, G.J.
Schmitz, Phys. Rev. B 70, 144514 (2004).

\bibitem{bean} C.P. Bean, Phys. Rev. Lett. 8, 250 (1962).

\bibitem{frietz} W.A. Frietz, W.W. Webb, Phys. Rev. 178, 657 (1969).

\bibitem{vkh} V.V. Val'kov, B.P. Khrustalev, JETF 80, 680 (1995) [Zh. Eksp. Teor. Fiz.
107, 1221 (1995)].

\bibitem{hilg} H. Hilgenkamp, J. Mannhart, Rev. Mod. Phys. 74, 485 (2002).

\bibitem{phC} M.I. Petrov, D.M. Gokhfeld, D.A. Balaev, K.A. Shaihutdinov, R.
K\"{u}mmel, Physica C 408, 620 (2004).

\bibitem{arxiv} D.M. Gokhfeld, D.A. Balaev, K.A. Shaykhutdinov, S.I. Popkov, M.I. Petrov,
cond\D mat/0410112 (2004).

\end{thebibliography}
\end{document}